# Temporal rhythms of uncertainty: Evaluating Mars claims in the Australian colonial press

**Richard de Grijs**
*Macquarie University, Sydney, Australia*

**Abstract.** Favourable oppositions of Mars gave nineteenth-century astronomers recurring opportunities to test claims about the planet, while telegraphic news and newspaper reprinting carried those claims well beyond the professional observatories. Historians have established the importance of these events to the development of mass-media astronomy. Here, I examine a related question in the Australian colonial press: how did the standing of particular claims change as brief reports, local responses and fuller accounts appeared at different times? Drawing on colonial newspaper coverage from 1875 to 1899, I follow episodes surrounding the major Mars oppositions of 1892 and 1894. Reports of observations attributed to "Professor Pickering" in 1892 prompted both rapid circulation and immediate scrutiny. In 1894, a cable about William W. Campbell's spectroscopic work provoked local disagreement before more detailed accounts became available. These sequences show that a Mars opposition concentrated attention without imposing a uniform progression from anticipation to discovery and subsequent caution. Newspapers carried assessments and renewed speculation at the same time. The Australian case extends work on event astronomy by showing how the timing and form of information arriving in a colonial press network affected the public evaluation of astronomical claims.

## 1. Introduction

The planet Mars was a recurring news event in late nineteenth-century astronomy. Its favourable "oppositions" brought concentrated observing campaigns, while telegraphs and newspapers carried reports of sometimes disputed observations across long distances. Nall (2017, 2019) has described this relationship between astronomical events and mass media as "event astronomy," showing how the significance of claims developed through their movement across places and through time. In this article, I take that account as my starting point. I ask what happened when reports originating in the principal centres of planetary astronomy, located in the United States, Britain and Europe, entered the newspapers of the Australian colonies.

The key question I aim to address concerns the evaluation of individual claims as time progressed. An initial cable might offer a striking conclusion but with little supporting detail; a local astronomer might challenge its implications; a fuller report might arrive weeks or even months later. Meanwhile, newspapers could continue to print other, more expansive accounts. Following that sequence shows how a claim's apparent force depended partly on which version of it was available at a given moment. It also allows us to examine a colonial setting in which local observers could respond to overseas news and, on occasion, contribute observations of their own.

I will use the newspaper collection examined in my three earlier studies, but I ask a different question of it. My first study (de Grijs 2026a) addressed the interpretation of Mars and the authority claimed by different observers. The second (de Grijs 2026b) mapped the circulation, attribution and changing volume of reports, including the bursts associated with oppositions and their longer reprint tails. The third (de Grijs 2026c) examined humour as a way of responding to astronomical claims. Here, I follow dated sequences of claim and response, expressly focused on 1892 and 1894, to examine when qualification became available and whether it displaced the initial report. This article therefore builds on the circulation patterns established in de Grijs (2026b) without treating those patterns as a new discovery.

My source base consists of Australian newspaper items collected for those studies, read here in publication order. I compare the wording and attribution of short cables, editorial responses, locally authored astronomical commentary and later, more comprehensive reports.

Reprints help establish the reach and persistence of a formulation, although multiple appearances of the same text do not constitute independent assessments. Newspaper reports show what arguments were available in print; they do not establish what any particular reader may have believed.

"Colonial" in the context of this paper refers to the Australian settler colonies before Federation in 1901. Their newspapers participated in imperial and wider international news networks through telegraphy, correspondence and reprinting (Potter 2003; Secord 2004). Their distance from the main observatories made the order in which information arrived especially consequential, whereas Australian astronomers and editors also supplied their own assessments.

## 2. Planetary oppositions as epistemic events

Throughout the nineteenth century, favourable oppositions of Mars were widely recognised as periods of intensified observational opportunity. In astronomical terms, an opposition occurs when a planet lies opposite the Sun in the sky as seen from Earth, such that it rises at sunset, remains visible throughout the night and reaches its maximum apparent brightness and size at that time. For Mars, oppositions recur at approximately two-year intervals, although their observational quality varies depending on the planet's orbital position.

Although this definition describes a geometric alignment, its historical significance extends beyond celestial mechanics. Oppositions functioned as temporally bounded observational regimes, that is, predictable intervals when astronomers expected enhanced visibility and therefore concentrated their observational efforts. In this sense, an opposition was not merely a celestial event but a coordinated moment in scientific practice, defining when observations were undertaken, compared and publicly reported.

The predictable return of Mars to favourable viewing conditions supplied both astronomers and journalists with a timetable. Nall's account of event astronomy shows that such occasions generated public, contested claims whose meanings changed as news moved through the mass media (Nall 2017, 2019). Australian reporting lets us follow another part of that movement: the interval between receiving a striking overseas claim and obtaining either criticism or fuller information about it. The interval varied from case to case. A response could appear almost immediately, while a more detailed account might take months to arrive.

I use "temporal rhythms" to describe these observable relationships among anticipation, first report, response and later recirculation. They are analytical questions, not fixed stages through which every opposition must pass. In particular, the appearance of a cautious assessment did not necessarily end the circulation of more speculative claims. The following sections therefore attend to dated newspaper sequences and to the coexistence of different assessments within them.

## 3. Why Mars oppositions mattered astronomically and journalistically

Planetary oppositions mattered not only because they improved observational conditions but because they reorganised the relationship between observation, interpretation and communication. Their significance lay less in their technical properties alone than in their capacity to synchronise scientific and media practices across dispersed locations.

For Mars, oppositions created periods when the planet appeared brighter, larger and more detailed than at other times. Nineteenth-century astronomers regarded these intervals as critical opportunities to examine surface features that were otherwise faint or ambiguous. Because such observations were often uncertain and required repeated verification, favourable oppositions were considered periods when existing interpretations might be confirmed, refined or challenged:

> The apparent disk [of Mars] is now larger in the proportion of three to one than when the planet is in aphelion [farthest distance], while the illumination is more brilliant in the proportion of three to two. At the same time the planet is nearer perihelion [closest distance] than previously for more than thirty years; so that in the heaven its brightness is but little inferior to that of Jupiter. (*Sydney Mail and New South Wales Advertiser*, 27 October 1877; and syndication)

These observational rhythms were widely communicated beyond specialist circles. Newspapers translated astronomical conditions into accessible terms, emphasising that Mars would soon appear unusually prominent in the night sky. In doing so, they framed oppositions as episodes of heightened epistemic possibility, positioning them as occasions when new knowledge might emerge. Reports describing the increased brightness or apparent size of the planet did more than convey technical information; they established an expectation that forthcoming observations would carry particular authority. As one newspaper later reflected:

> Although the superficial condition of Mars has been a matter of research since the days of [William] Herschel [1738–1822], and, for the matter of that, since Galileo [Galilei; 1564–1642], it was not until the "opposition" in 1887 [*sic*; 1877] that any accurate maps of its surface were forthcoming. These maps were confirmed and improved, in 1882, when next again the planet was nearest to us; and this year [1892] they will once more be checked and enlarged by the assistance of instruments of which observers of former years had not the advantage. (*Geelong Advertiser*, 17 September 1892)

The influential observations made during the 1877 opposition by Giovanni Schiaparelli (1835–1910) intensified this dynamic (Sheehan 1996). His reports of linear features—*canali*, later widely described as "canals"—introduced an observational problem that subsequent oppositions promised to resolve. Each favourable alignment thus became an implicit test of earlier claims, transforming cyclical celestial motion into a sequence of observational trials. Oppositions were no longer simply opportunities to observe; they became moments against which the credibility of prior interpretations could be assessed.

The predictability of these events also made them particularly compatible with journalistic reporting. Because oppositions could be calculated in advance, newspapers constructed extended temporal arcs: announcing forthcoming observational opportunities, reporting results as they emerged and retrospectively evaluating their significance. In this sense, oppositions functioned as media events, aligning the temporal rhythms of journalism with those of celestial mechanics.

This narrative structure was reinforced by the transnational circulation of scientific information within imperial news networks (Potter 2003; Secord 2004; Lightman 2007). Colonial newspapers drew extensively on telegraphic reports, overseas correspondence and reprinted material from European and American sources. Observations made at distant observatories were rapidly incorporated into local reporting, enabling geographically dispersed audiences to engage with the same sequence of events in near-synchronous fashion.

Newspapers also connected Martian markings to questions about water, cultivation and possible inhabitants. Such interpretations travelled internationally, but their appearance in Australian papers placed them before readers for whom irrigation and water scarcity were familiar concerns. The following account, printed in the *Bendigo Advertiser*, shows how the press presented the proposed canals as part of a wider argument about life on Mars:

> We find, in the first place, that the broad physical conditions of the planet are not antagonistic to some form of life; secondly, that there is an apparent dearth of water upon the planet's surface, and therefore, if beings of sufficient intelligence inhabited it, they would have to resort to irrigation to support life; thirdly, that there turns out to be a network of markings covering the disc precisely counterparting what a system of irrigation would look like and, lastly, that there is a set of spots placed where we should

> expect to find the lands thus artificially fertilised, and behaving as such constructed oases should. (*Bendigo Advertiser*, 5 October 1895)

Lane (2011) examines how geographical representations and comparisons with terrestrial landscapes helped make claims about Mars persuasive, whereas Markley (2005) places connections between astronomy and the cultural imagination within a longer history of writing about Mars. The *Bendigo Advertiser* passage demonstrates the circulation of this interpretation in Australia; the extract alone does not establish that it originated there (cf. de Grijs 2026a).

Planetary oppositions therefore linked multiple scales of activity. Orbital motion created periodic moments of enhanced visibility, astronomers organised their observational practices around these cycles, newspapers transformed them into narrative events and readers interpreted them through culturally situated frameworks. In doing so, oppositions structured not only when observations occurred but how their significance was understood within the public sphere.

## 4. Anticipating discovery: The approach of opposition in the colonial press

Australian newspapers could identify the questions an approaching opposition was expected to address. On 20 July 1892, the Adelaide *Advertiser* noted that Mars would reach opposition on 4 August and that observers would have a particularly favourable opportunity to examine its supposed canals. The notice drew attention both to the improved viewing conditions and to the unresolved interpretation of the markings:

> Special interest will attach to the examination of the long narrow markings termed "canals," which are visible on the planet, and which it has been suggested by some observers are engineering works on a cyclopean scale. (*Advertiser*, 20 July 1892, p. 4)

This was an expectation of closer examination, rather than a prediction that the canal question would be settled. The approaching opposition gave newspapers a reason to return to an existing claim before the new reports arrived.

In May 1894, anticipation took the form of a request for observations rather than a general promise of discovery. John Tebbutt (1834–1916), the Australian astronomer and meteorologist, relayed Professor Pickering's invitation to Australian observers to examine features seen in 1892 when Mars returned to a comparable point in its seasonal cycle. A subsequent notice by Henry Chamberlain Russell (1836–1907), the New South Wales Government astronomer, identified the correspondent as William Henry Pickering (1858–1938) and asked suitably equipped local observers to assist (Tebbutt 1894; *Sydney Morning Herald*, 24 May 1894; and frequent syndication). This proposed comparison took place months before the October 1894 opposition. It shows that the observational timetable was shaped by the feature under investigation as well as by the date of closest viewing. It also placed Australian observers within the proposed test of an overseas claim. Russell wrote,

> It will be remembered that in May last I called attention to a circular from Professor W. H. Pickering, of United States, asking for special observations of the Y mark on the planet Mars. Although a careful watch was kept on the planet here, I failed to see anything of the marking in question, and, so far as I am aware, no one else in the colony was able to see anything of it. (*Sydney Morning Herald*, 24 August 1894, p. 6)

Russell's report supplied an outcome to the proposed comparison: neither he nor, according to Russell's account, other observers in the colony had recovered the marking. The result was reported in August, before the October opposition. The question raised by the 1892 observation was therefore being tested on a timetable determined by the feature's expected seasonal appearance, not solely by the date of opposition.

## 5. Claims and responses around the 1892 and 1894 oppositions

Nall reconstructs William H. Pickering's collaboration with the *New York Herald* and the telegraphic reporting from Peru that helped produce the "great Mars boom" of 1892 (Nall 2017; Nall 2019: chap. 3). Harvard's earlier Boyden expeditions had already exposed the organisational importance of communication between remote observing sites and the home observatory (Nisbett Becker 2009). The Australian reports belonged to this wider news event. My concern here is with their subsequent presentation and assessment in Australia: the identity attached to a brief announcement, the speed with which a local astronomer challenged it, and the different versions that remained in circulation.

The 1892 reports attributed to Professor Pickering show how quickly a claim and its scrutiny could enter Australian newspapers. Early September cables announced the discovery of eleven lakes and two mountain ranges. Some reports named E. C. (Edward Charles) Pickering (1846–1919), the director of the Harvard College Observatory, although the Mars observer concerned was his brother, William H. Pickering (*Tasmanian News*, 3 September 1892; *Argus*, 5 September 1892; and frequent syndication). The attribution matters because attaching a familiar institutional identity to a brief cable affected how the report was presented. The surviving Australian items establish the misidentification in print; they do not establish where it first arose.

The initial claim did not circulate without challenge. The *Argus* printed the cable with comments from the Victorian Government astronomer, Robert Lewis Ellery (1827–1908), who questioned what could be inferred from such a short account (*Argus*, 5 September 1892):

> Putting all these things together, and the statement in the cable message, and knowing that Professor Pickering had no instruments there such as are in the Northern Hemisphere, Mr Ellery thought that when the facts became known it would be found that some bare statement of Professor Pickering's had been enlarged upon. (*Argus*, 5 September 1892)

Ellery's objection about the available instruments requires some caution. The same report identified the observer as "E. C." Pickering, whereas his brother William was engaged in obtaining the actual visual observations at Arequipa. The passage demonstrates immediate local scrutiny of the cable, but it cannot be treated as a fully informed assessment of William's equipment (Nall 2017).

Further discussion followed within days. Thus, the sequence cannot be described simply as an initial period of unqualified acceptance followed by a later correction. A dramatic claim and a local expert's reservations were available simultaneously near the beginning of its Australian circulation. Subsequent versions and commentary prolonged the episode, but they did not give every newspaper item the same degree of certainty.

The 1894 dispute over the Martian atmosphere reveals a different timing. In August, the *Sydney Morning Herald* relayed news of William Wallace Campbell's (1862–1938) spectroscopic findings at Lick Observatory (Mt. Hamilton, California). Its accompanying commentary presented the implications in categorical terms (*Sydney Morning Herald*, 20 August 1894; and syndication):

> Professor CAMPBELL, of the Lick Observatory, in California, asserts that the planet Mars is destitute of an atmosphere. Assuming that further investigation should maintain this assertion, then total abandonment or essential modification awaits the interesting theories which during many years have

> been formed about this member of our system. To prove that the planet has no atmosphere is to prove that it does not contain life in any sense conceivable by us. (*Sydney Morning Herald*, 20 August 1894)

The next day, Ellery challenged that conclusion by appealing to other observations, including the Martian polar caps (*Argus*, 21 August 1894; and syndication). Russell urged readers to wait for the details of Campbell's investigation before judging the result (*Sydney Morning Herald*, 24 August 1894, p. 4; and syndication):

> Observers of the surface of Mars are familiar with the strange variations in the visibility of the markings of the planet, and naturally they have assumed that this uncertainty is caused by clouds of vapour floating in the atmosphere of the planet. This has been the accepted view for many years, but it is difficult to say what may have been done at the Lick Observatory. They have there a splendid opportunity and one of the largest and best furnished telescopes in the world, and they may have been able to ascertain some facts which Dr. Huggins could not see with his smaller instrument. I am not in a position to express an opinion as to what has been gleaned at the Lick Observatory until I see a report in full. (*Sydney Morning Herald*, 24 August 1894, p. 4)

Local astronomers therefore did more than repeat or reject a metropolitan announcement: they distinguished a reported result from the stronger conclusion drawn from it.

The distinction became clearer when fuller accounts circulated. Campbell's published paper found no spectroscopic evidence of Martian water vapour or an extensive atmosphere. He expressly stated that his observations did not prove Mars had no atmosphere; he considered the polar caps evidence for a thin atmosphere and water vapour (Campbell 1894). By December, an Australian newspaper reported that the more detailed American account supplied no greater spectroscopic evidence for a Martian atmosphere than did the Moon's spectrum, while allowing the possibility of a very tenuous one (*Maitland Daily Mercury*, 12 December 1894). The sequence runs from a compressed cable, through prompt local objections, to a qualified technical account that reached readers later. Its interest lies in the different claims those forms of reporting made available, not in a single change of collective opinion.

## 6. Later accounts and continuing disagreement

The arrival of later information did not produce a uniform retreat from speculation. The December account of Campbell's result supplied qualifications missing from the August cable, yet other late 1894 reports continued to discuss Martian canals, surface changes and possible inhabitants. Lowell's interpretation of a geometrically organised canal system appeared in the *Capricornian* on 29 December 1894, while an account attributed to Schiaparelli in the *Evening News* on 26 December stressed the limits of what observation could establish. These appeared in separate newspapers. Their value as evidence lies in their coexistence across the press record, not in an editorial juxtaposition within one article.

Reprinting also gave claims different lifespans. A short announcement could travel rapidly and retain its categorical wording; a qualified explanation could arrive later and circulate through different papers. Readers of one newspaper or one issue need not have encountered both. The evidence therefore supports a more limited conclusion than a general movement from excitement to moderation: Australian newspapers made differing degrees of confidence available at overlapping times.

Nor did a single opposition settle the questions it raised. An *Argus* account in November 1895 described a theory of Mars as upheld by some and denounced by others (*Argus*, 16 November 1895). Such disagreement kept earlier observations available for reconsideration at later viewing opportunities. The recurring opposition supplied a reason to reopen a question, while the newspaper's own schedules of cables, correspondence and reprints determined when particular answers appeared.

Alternative explanations continued to appear well after the 1894 opposition. In March 1897, Sydney's *Daily Telegraph* reported Marius Tepperberg's suggestion that snowfall could explain the canals and their apparent doubling:

> Yet another view of the canals of Mars has been propounded. The opinion that the appearances presented by our [E]arth's neighbo[u]r among the planets were due to vegetable growth was duly noted in this column some months ago. Now Herr M[arius]. Tepperberg, of the Hague, maintains that the canal systems represent snow-falls, and that their periodical doubling is to be explained on optical grounds connected with the mode of deposition of the snow. This view of matters, however, does not appear to find favo[u]r in the eyes of astronomers, who regard the vegetation theory as a more feasible explanation. (*Daily Telegraph*, 27 March 1897)

The paper also noted that astronomers favoured a competing explanation based on vegetation. The item shows the continued circulation of rival interpretations, rather than a final resolution following an observational campaign.

## 7. Rhythms of scientific credibility

Nall's (2017, 2019) accounts of event astronomy establish that the oppositions of Mars and the mass media together generated recurrent occasions for disputed astronomical news. The Australian cases show how the evaluation of that news worked within a geographically distant press network. In 1892, a striking report attributed to Pickering circulated alongside rapid local scrutiny. In 1894, local astronomers challenged a categorical account of Campbell's work before a fuller, more qualified explanation arrived. These sequences make the timing of information central to the history of a claim.

The analysis also qualifies any simple cycle of anticipation, proliferation and cooling. Astronomers could organise a comparative campaign before the date of opposition; objections could appear alongside an initial announcement; and speculative reports could continue after more cautious accounts became available. The recurrence of viewing opportunities mattered, but neither celestial mechanics nor journalistic practice imposed the same order of assessment on every episode.

These findings extend the established history of Mars and the mass media into the Australian colonial press. Newspapers connected distant observatories to local commentators, and their differing speeds of transmission and forms of presentation affected which version of a claim appeared at a particular time. The result was a public record in which uncertainty could be examined, contested and renewed without reaching a common resolution.

**Competing interests**: The author declares that there are no competing financial or non-financial interests related to this research.